\documentstyle[12pt]{article}
\begin{document}
\newcommand{\DUR}{\em Department of Physics\\
                      University of Durham \\
                      Science Laboratories, South Road\\
                      Durham, DH1 3LE, U.K.}

\def\lhs{{\it l.h.s.\/ }}
\def\rhs{{\it r.h.s.\/ }}
\def\GeV{{\rm GeV}}
\def\bqn{\begin{equation}}
\def\eqn{\end{equation}}
\def\bqna{\begin{eqnarray}}
\def\eqna{\end{eqnarray}}
\def\nn{\nonumber}
\def\bit{\bibitem}
\def\vs{\vspace{.25in}}
\def\thefootnote{\fnsymbol{footnote}}

\def\QQa{\renewcommand{\baselinestretch}{1.3}\Huge\large\normalsize}
\def\gl{\tilde{g}}
\def\mg{m_{\gl}}
\def\as{\alpha_s}
\def\asmz{\alpha_s(M_Z)}
\def\GeV{{\rm GeV}}
\def\pt{P_T^{\rm min}}
\def\deltar{\Delta r^{\rm min}}
\def\etamax{|\eta|_{\rm max}}
\def\ycut{y_{\rm cut}}
\def\thetanr{\theta^*_{\mbox{\sc nr}}}
\def\degree{^\circ}
\begin{titlepage}
\vspace*{-1cm}
\begin{flushright}
DTP/94/50   \\
June 1994 \\
\end{flushright}
\vskip 1.cm
\begin{center}
{\Large\bf
Angular Correlations and Light Gluinos  in    \\[2mm]
Multi-jet Photoproduction at HERA}
\vskip 1.cm
{\large R.~Mu\~noz-Tapia}
\vskip .2cm
{\it Department of Physics, University of Durham \\
Durham DH1 3LE, England }\\
\vskip   .4cm
and
\vskip .4cm
{\large  W.J.~Stirling}
\vskip .2cm
{\it Departments of Physics and Mathematical Sciences, University of Durham \\
Durham DH1 3LE, England }\\
\vskip 1cm
\end{center}
\begin{abstract}
A  study of $3+1$ jet event photoproduction at HERA is
presented.
We define an angular variable which  is sensitive to the topology
of the final state jets and  is  therefore  able to discriminate between the
different contributing subprocesses, and between QCD and an abelian
gluon model.
We also investigate the contribution from the
direct production of light gluinos to the $3+1$ cross section.
\end{abstract}
\vfill
\end{titlepage}
\newpage

\QQa
\section{Introduction}

The HERA electron-proton collider  provides a unique opportunity
to study photoproduction processes  at high energy \cite{schuler}.
When the scattering
angle of the electron is small,  the square of the four-momentum
transfer is also small and the exchanged particle can be
considered as a quasi-real photon. Such photons interact electromagnetically
with leptons and quarks, but also have a hadronic component ({\it i.e.} quark
and gluon constituents)  from branching processes such as
$\gamma\to q \bar q,\ \gamma\to q \bar q g$ etc. These two types of photon
interactions give rise to what are called `direct' and `resolved' processes
respectively.
Calculations indicate that  for the paradigm  hard
photoproduction process,  the
production of two jets, the resolved part gives an important contribution
at low and moderate jet transverse momenta  \cite{sti-kun,exp}.

The centre-of-mass energy of  photon-proton scattering at HERA -- typically of
order 200~GeV -- is large  enough to allow the production of multijet
events, {\it i.e.} events with more than two large $p_T$ jets in the
final state. As at the LEP $e^+e^-$ collider, these can
provide detailed tests of QCD matrix elements, as well as measurements
of the strong coupling $\as$ from comparing cross sections for the production
of different numbers of jets.

In this study, we  perform a detailed analysis of  the production of
$3+1$  large $p_T$ jet events at HERA.\footnote{The `$+1$' refers to the
proton remnant jet.}
We will be interested in the case when all the jet transverse momenta are
large, in which case the subprocess centre-of-mass energy is a sizeable
fraction of the overall $\gamma p$ energy and as a result
the resolved part of the photon gives a negligible contribution.
We therefore restrict ourselves to  the subprocesses $\gamma q, \gamma g
\to (q \bar q, qgg, q \bar q g, \ldots)$.
The idea is to try to perform the same type of QCD matrix-element
tests  that have
been performed using  the four-jet sample of jet events at LEP, where
the dominant processes are $\gamma^* \to q \bar q gg,\ qq \bar q \bar q$.

As at  LEP, it should be feasible to define angular
distributions that help in discriminating the parton composition of
the final states. At LEP, these have been used, for example, to
distinguish
QCD from an abelian gluon theory  and also to put bounds on the
existence
of new light fermion species. This is particularly important since
light, neutral, coloured fermions, the so-called `light gluinos', have not yet
been conclusively ruled out by experiments \cite{gluino-window}.
There has been some speculation that the
contribution of such particles to the QCD $\beta$-function
 would reconcile the measurements of
$\as$ at low and high energies \cite{ga1,ga2,ga3,ga4}.
Recently, several studies suggesting methods of closing the existing window
have been published \cite{gc1,gc2,gc3,gc4}.
At the HERA collider, the $3+1$ cross section is the leading-order
cross section for the pair production  of such  gluinos, and might
therefore provide  the first direct evidence for their existence.

The $2\to 3$ matrix elements which we use to calculate the $3+1$
jet cross sections give rise, in general,
 to infra-red and collinear singularities
when the final state particles are soft and/or collinear with each other
and with the incoming particles. We will regulate these in the standard
way
by requiring that the transverse momenta of the final-state particles (jets)
 exceeds a certain cutoff $\pt$, and by requiring a minimum
$\Delta r=(\Delta \eta^2 + \Delta \phi^2)^{1/2}$ separation in the
pseudorapidity-azimuthal plane $(\eta,\phi)$  \cite{jet-al}.
This way of defining jet final states has already been used successfully
at HERA \cite{jets}.
The maximum pseudorapidity of each jet is also restricted,
to keep the jets away from the beam direction.

In the following section we describe the general features of the $3+1$
jet cross section, and calculate the contributions to the total cross
section from the different subprocesses. In Section 3 we introduce an
angle which characterizes the topology of the final state and which
can, in principle, discriminate between the different subprocesses.
In Section 4 we discuss  a possible `light gluino' contribution
to the cross section, and in Section 5 we present our conclusions.

\section{The total $3+1$ jet cross section}

The total cross section for a $3+1$ jets final state can be written
schematically as
\bqn
\sigma (e p \longrightarrow e + 3 {\rm jets} + X)=
\sum_{a,c_i=q,g} G_{\gamma/e}\;  * \;  G_{a/p} \; * \;
\hat{\sigma}(\gamma a \rightarrow c_1 c_2 c_3),
\eqn
where $G_{\gamma/e}$ and $G_{a/p}$ denote the photon content of the electron
and the parton content of the proton respectively .
For the latter, we use the
MRS(D$_0'$) set \cite{MRS}, although none the quantities  that we will
calculate
will be particularly sensitive to this choice -- our quarks and gluons
will be probed at relatively large $x$ where they are well constrained
by deep inelastic and other data.
The symbol $*$ denotes a convolution
operation and $\hat{\sigma}$ refers to the partonic cross sections of the
relevant processes.
The $Q^2$ scale in the parton
distributions and in the strong coupling  constants in the subprocess cross
sections is set equal to
the minimum jet transverse momentum in each event.
To a very high accuracy the Weizs\"{a}cker-Williams approximation \cite{WW}
(already
implicit in Eq.~(1)) can be used for the photon content of the electron
\cite{zerwas}
\bqn
G_{\gamma/e}=\frac{\alpha}{2\pi}\log\left( \frac{s}{4m^2} \right)
\frac{1+(1-x)^2}{x},
\eqn
where $\alpha$ is the electromagnetic coupling constant, $s$  is the
centre-of-mass energy squared, and $x$ is the fraction of energy lost by the
electron $x=(E-E^{'})/E$.
In a recent paper Frixione {\it et al.} \cite{frixione} have
studied  the validity of the
approximation (see also \cite{lebedev}). The sub-leading corrections are
 negative, indicating that  the above approximation
always overestimates the cross section. None of our results, however,
depend sensitively on the absolute size of our cross sections.

In Fig.~1 we show the total cross section for the process
$e + p \rightarrow e + {\rm 3\; jets} + X$ as a function of the minimum
transverse momentum cut $\pt$. The contributions of the
different subprocesses are also
shown. We have set the cut-off $\deltar$ of the jet-defining algorithm to
1.0, in accordance with Ref.~\cite{snowmass},
 and the maximum rapidity of the jets (in the $\gamma p$
centre-of-mass frame) is 
2.0.
 Notice that the 2 quark $+$ 1 gluon configuration dominates at low $\pt$.
This contribution is proportional to the gluon distribution in the
proton, which at small $x$ ({\it i.e.} small $\pt$) is larger
than the quark distributions.
At large $\pt$, on the other hand, the partons are probed at large $x$
and the quark-induced subprocesses dominate.
The crossing between the two  occurs at around $\pt =  25\ \GeV/c$.

\section{Angular distributions}

Angular variables for multijet final states
 have long been  used in $e^+ e^-$
colliders, see for example Ref.~\cite{hebbeker}.
Due to their different helicity and colour properties, quarks and gluons
exhibit different behaviour in certain
kinematic variables.  This fact can be used to
discriminate the parton content of the final-state jets.
For example,  for  $e^+ e^-\longrightarrow$ 4 jets
the modified Nachtmann-Reiter
angle and the azimuthal angle between the planes defined by the
two final-state jet pairs have led to important
tests of the QCD structure of the matrix elements \cite{angles}.

Here we attempt to define an analogous angular variable which
is suited to the study of  $3+1$ jet final states at the HERA
$ep$ collider.\footnote{We assume in what follows that a sufficiently
large sample of such events will be collected over the lifetime of the
machine in order to perform such  a study.}
First, the jets are ordered
according to their transverse momentum $P_T$. Then  the
angle $\theta_H$ between the planes formed by the highest $P_T$ jet and the
beam and the plane formed by the other two jets is computed.
In addition, we  require the highest $P_T$ jet to be  central in
rapidity,
so that the typical configuration is that of an energetic jet in one
hemisphere balanced by two less energetic jets in the opposite
hemisphere.

One important difference between LEP and HERA is that only for the
former do the lab and multijet centre-of-mass frames coincide.
Any information about the angular correlations between the final-state
partons at HERA tends to be smeared out in going from the parton subprocess
frame to the lab frame.
In order to understand the underlying QCD physics, therefore, we first
analyse the angular distribution in the subprocess centre-of-mass frame.

We first consider the angular distributions between the final states
{\it without} a central
rapidity cut on the highest $P_T$ jet.
In this case we find that the 1 quark $+$ 2 gluons final state
($\gamma q \to q gg$)
 shows a qualitatively different behaviour  to the other
two processes. For  this process we would  expect  the quark to
 be the most energetic
particle because of the infra-red singularities for soft gluon emission.
(This is analogous to the $q \bar q gg$ final state at LEP where the
gluons are generally the softest jets.)
The distribution
in the polar angle formed by the beam  direction and the quark jet
is fairly flat, having a broad peak around $90\degree$ and decreasing at
$0\degree$ and $180\degree$ due to the rapidity and $\pt$ cuts.
However in the 3 quarks final state ($\gamma q \to q q \bar q$),
the  highest $P_T$ jet has
a distribution of the angle with the beam direction peaked at low angles.
This difference can be understood as follows. In the first case,
diagrams involving the trilinear gluon coupling give an important
contribution and the subprocess scattering is effectively
$\gamma q \to q g^* (\to gg)$, with $t$-channel fermion exchange at
small angles. In the second case, the dominant configuration
is effectively $\gamma \to q \bar q$ followed by $qq \to qq$, which
proceeds via $t$-channel gluon exchange and is therefore more peaked at
small angles.
The 2 quark+1 gluon final state process ($\gamma g\to q \bar q g$)
 shows a distribution in polar angle of
the largest $P_T$ jet similar to that of the 3 quark final state.
Here, again,  the dominant configuration involves an initial
state $\gamma \to q \bar q$ splitting followed by $q g \to q g$
involving $t$-channel gluon exchange.

Now since we are interested in increasing the sensitivity to the triple-gluon
vertex in {\it final-state} gluon radiation, as at LEP,
it is sensible to require  the fastest $P_T$ jet to
be central in rapidity, thus suppressing
processes involving {\it initial-state} splitting, like
 the second and third type  discussed above.

In Fig.~2 we show the total cross section for the 3+1 process in the
HERA frame when the central cut $\vert\eta\vert < 0.5$ for the largest
$P_T$ jet  is included.
Note that the overall decrease in rate compared to Fig.~1 is
not particularly significant.

In Fig.~3 the distributions in the angle ${\theta}_H$  -- in the
subprocess centre-of-mass frame -- are presented for the
sum of all processes,  and for each subprocess normalized separately.
As we had anticipated,  there is a distinct difference  in
the distributions depending
on the composition of the final-state  jets. The planes
corresponding to the
1 quark $+$ 2 gluon final state tend to be aligned perpendicularly
($\theta_H \sim 90\degree$) while
the other two processes show a peak at low angles.
This can be understood as follows.
When the beam   direction   is coplanar with
 the 3 quark (or 2 quark $+$ 1 gluon)
final state, the important $t$-channel gluon exchange contribution
is maximized, thus favouring low $\theta_H$.
However, in  the  1 quark $+$ 2 gluon final state case the pole
structure is milder and other kinematic effects come into play.
The two gluons are expected to end up as  the softest jets,
whereupon they
define one of the planes for computing $\theta_H$. For these two
jets, the   rapidity and $P_T$ cuts have a bigger impact, and the number
of events with the beam direction perpendicular to the plane
of the final-state  jets is enhanced.

In order to distinguish   kinematic and dynamical  effects,
it is useful to compare the QCD angular distributions with those
of a  phase-space model,
 where the matrix elements are constant. In Fig.~4 the phase
space  $\theta_H$ distribution is compared with the distribution
for the QCD 1 quark $+$ 2 gluons final state, and also with that
of an `abelian' QCD model \cite{abelian}, {\it i.e.}
a U(1)$_3$ gauge theory with a coupling constant
 $\alpha_{ABE}=4/3 \alpha_s$ chosen
to compensate the QCD $q \to q g$ colour factor.
This model provides a useful benchmark for  demonstrating sensitivity
to the triple-gluon vertex.
We see from  Fig.~4 that the abelian and phase-space distributions have
roughly the same shape,
both showing a small peak at low angles and
a decrease at higher angles, in contrast to the QCD result which is
peaked at high angles.
The decrease of the phase space distribution is due to the angular
dependence introduced by the $P_T$ ordering -- without this ordering the
distribution would be essentially flat. Notice that the configuration
of the final jets produced in the phase-space model is
different to that of the  QCD 1 quark $+$ 2 gluons subprocess,
where, on average, two of
the jets are significantly softer than the third. This explains why the
 cuts  have a different effect in the two cases.
 This distinctive behaviour gradually disappears as the $P_T^{\rm min}$
cut is increased above $20\ \GeV $, when most the events at
low $\theta_H$ are removed.

We next consider the $\theta_H$ distribution in the HERA lab frame.
The boost induced by the
more energetic proton beam squeezes the difference between distributions
into a smaller part of the angular range. In addition, there is a greater
sensitivity to the jet rapidity cuts.\footnote{To compensate for
this, we increase $\etamax$ from 2.0 to 3.0 in what follows.}
Figure~5 shows the same distributions in Fig.~3 but now in the HERA
frame. Again, we see that the distribution for the  1 quark $+$ 2 gluons
 final state is
larger  for higher angles, while the distributions for
the other processes decrease at higher angles.
Although the effect is less pronounced than in the centre-of-mass
frame, the distributions still show differences of  order 50\% at
perpendicular angles. The main problem  here
is that the pattern of each curve
starts to be distinctive only from about  $60\degree$ onwards.
The comparison  of  the QCD 1 quark $+$ 2 gluons process
with the abelian and phase-space  models is shown in Fig.~6.
The differences in shape
again only start to be noticeable at higher angles,
because most of the events at low angles have been boosted
out of the rapidity  acceptance.

\section{Light gluino production}

The production  of $3+1$ jets at HERA is the leading process
for the pair production of gluinos: $\gamma q \to q \gl \gl $.
In a previous study \cite{previous}, we analysed the angular
correlations in 4-jet production at LEP to investigate the effect
of light gluino pairs in the final state. Here we do the same for
$3+1$ jet production at HERA.
As an application of the shape distributions introduced in the
previous section, one can  study  the influence of
a light gluino particle.
Note that other methods of detecting light gluinos at HERA,
in particular through their effect on  deep inelastic structure
functions, have been shown to be very difficult
\cite{rs-gluino,deep-inelastic}.

In Fig.~7 we show the predicted total cross section for the photoproduction
of gluino pairs in $3+1$ jet events as a function of $\pt$, compared
with the QCD result. Since there is a difference of almost two
orders of magnitude between the two, it will be practically
impossible to detect any effect from the total cross section alone.

In Fig.~8  the $\theta_H$ distribution for gluino photoproduction
in the partonic centre-of-mass and  HERA frames is shown. Notice that the
shape is very similar to that of the 1 quark $+$ 2 gluon final state. The
Feynman diagrams
contributing to gluino-pair production coincide with those containing
the triple-gluon vertex in the $qgg$ case. The same arguments given in the
previous section apply here, and so  the plane formed by the light gluinos will
be preferentially oriented perpendicular to the direction of the beam.
However, the difference in shape is not big enough to compensate for the
overall smallness of
the gluino contribution, as shown in Fig.~8.  The only hope will be
to look for the decay signature of a colourless glueballino ($g \tilde{g}$)
formed after the hadronization process,
as suggested in Refs.~\cite{gc3,gc4}.
If enough events are produced, the angular
distribution could be used to perform a further test.

\section{Conclusions}

In this paper we have studied the photoproduction of
$3+1$ jet events at the HERA $ep$ collider.
An angular variable, defined in terms of the directions of the
final-state jets,
has been shown to discriminate between the different types of
contributing subprocess.
This  can be used to check the parton composition
of the final-state jets, to verify that QCD is favoured
over phase-space and abelian gluon models,
 and to put bounds on new light particle species.
In particular, we have studied the influence of a  light gluino,
whose existence is still controversial. Only a very small modification
on the angular distribution is expected.

\section*{Acknowledgements}
We are grateful to Tim Stelzer for discussions and for providing us
with the MADGRAPH
\cite{tim}
program for computing the matrix elements. Useful discussions
with  Valery Khoze are also gratefully acknowledged.
This research is supported in part by the Commission of the European
Communities `Human Capital and Mobility'  Network contract CHRX-CT92-0004
and (RMT) contract ERB4001GT921106.

\newcommand{\NPB}[3]{{ Nucl.\ Phys.\/} {\bf B#1} (19{#2}) {#3}}
\newcommand{\PLB}[3]{{ Phys.\ Lett.\/} {\bf #1B} (19{#2}) {#3}}
\newcommand{\PRD}[3]{{ Phys.\ Rev.\/}  {\bf D#1} (19{#2}) {#3}}
\newcommand{\PRE}[3]{{ Phys.\ Rep.\/}  {\bf #1} (19{#2}) {#3}}
\newcommand{\PRA}[3]{{ Phys.\ Rev.\/}  {\bf A#1} (19{#2}) {#3}}
\newcommand{\PRL}[3]{{ Phys.\ Rev.\ Lett.\/} {\bf #1} (19{#2}) {#3}}
\newcommand{\ZFP}[3]{{ Zeit.\ f.\ Phys.\/} {\bf C #1} (19{#2}) {#3}}
\newcommand{\IJA}[3]{{ Int.\ J.\ Mod.\ Phys.\/} {\bf A#1} (19{#2}) {#3}}

\newpage
\section*{Figure captions}

\begin{itemize}
\item[{[1]}]  Total cross section for the process $ e + p \rightarrow e +
               3 \ {\rm  jets} + X$ as a function of the minimum
              jet transverse momentum
              cut $\pt$. The UA1 jet  algorithm is used with
              $\deltar = 1.0$. The maximum  jet pseudorapidity is
              $\etamax  = 2.0$. The solid line corresponds to the
              total cross section, summed over all subprocesses.
              The dashed, dotted, and dash--dotted lines show
              the contributions from the   1 quark + 2 gluons,
               2 quarks + 1 gluon, and 3 quarks final states respectively.

\item[{[2]}]  Total cross section for the process $ e + p \rightarrow e + 3
              \ {\rm jets} + X$ as a function of the minimum jet
              transverse momentum
              cut $\pt$,  when a central cut $|\eta| \leq 0.5$ on the
              highest $P_T$ jet is included and $\etamax=3.0$.
              The solid line corresponds to the
              total cross section, summed over all subprocesses.
              The dashed, dotted, and dash--dotted lines show
              the contributions from the   1 quark + 2 gluons,
               2 quarks + 1 gluon, and 3 quarks final states respectively.

\item[{[3]}]  Shape distribution in the angular variable $\theta_H$,
             defined in the text, for the
             three-jet subprocesses in the centre-of-mass frame of the
             photon--parton subsystem.
             In the 3-quark final state the mass of
             the $b$ quark has been taken into account. The highest $P_T$ jet
             is required to be  central by imposing a
             cut pseudorapidity, $|\eta| \leq 0.5$.
             The solid line
             corresponds to the sum over all processes. Note that
             each line has been normalized separately to unit area.

\item[{[4]}] Shape distribution in the angular variable $\theta_H$
               for the 1 quark + 2 gluons QCD subprocess, and the
               same subprocess in an abelian model, in the photon--parton
               centre-of-mass frame.
               The dashed line
               shows the distribution corresponding to
               the phase-space model. A cut on the highest
               $P_T$ jet, $|\eta| \leq 0.5$, is also imposed.

\item[{[5]}] Shape distribution in the angular variable $\theta_H$
             for the three-jet subprocesses in the HERA lab frame.
             The maximum pseudorapidity is now $\etamax  = 3.0$.
             The dashed, dotted, and dash--dotted lines show the
             distributions for the 1 quark + 2 gluons, 2 quark + 1 gluon,
             and  3 quarks final states respectively. The solid line
             corresponds to the sum over all processes.

\item[{[6]}] Shape distribution in the angular variable $\theta_H$
              for the 1 quark + 2 gluons QCD subprocess, and the
              same subprocess in an abelian model, in the HERA lab frame.
              The dashed line
              shows the phase-space model distribution.

\item[{[7]}] The total cross section for the process $ e + p \rightarrow e +
              3\ {\rm jets} + X$  compared to $ e + p \rightarrow e + q
              + \gl \gl + X$ as a function of the minimum
              jet transverse momentum $\pt$, for $m_{\gl} = 5\ \GeV/c^2$.

\item[{[8]}]  Comparison of the $\theta_H$ shape distribution  for the total
               3-jet cross section and for the contribution from a light
              ($m_{\gl} = 5\ \GeV/c^2$) gluino. The dash-dotted,
               solid and dashed
              lines show the gluino, 3-jet QCD and combined distributions
              respectively (a) in the photon-parton centre-of-mass frame,
              and (b) in the HERA lab frame.

\end{itemize}

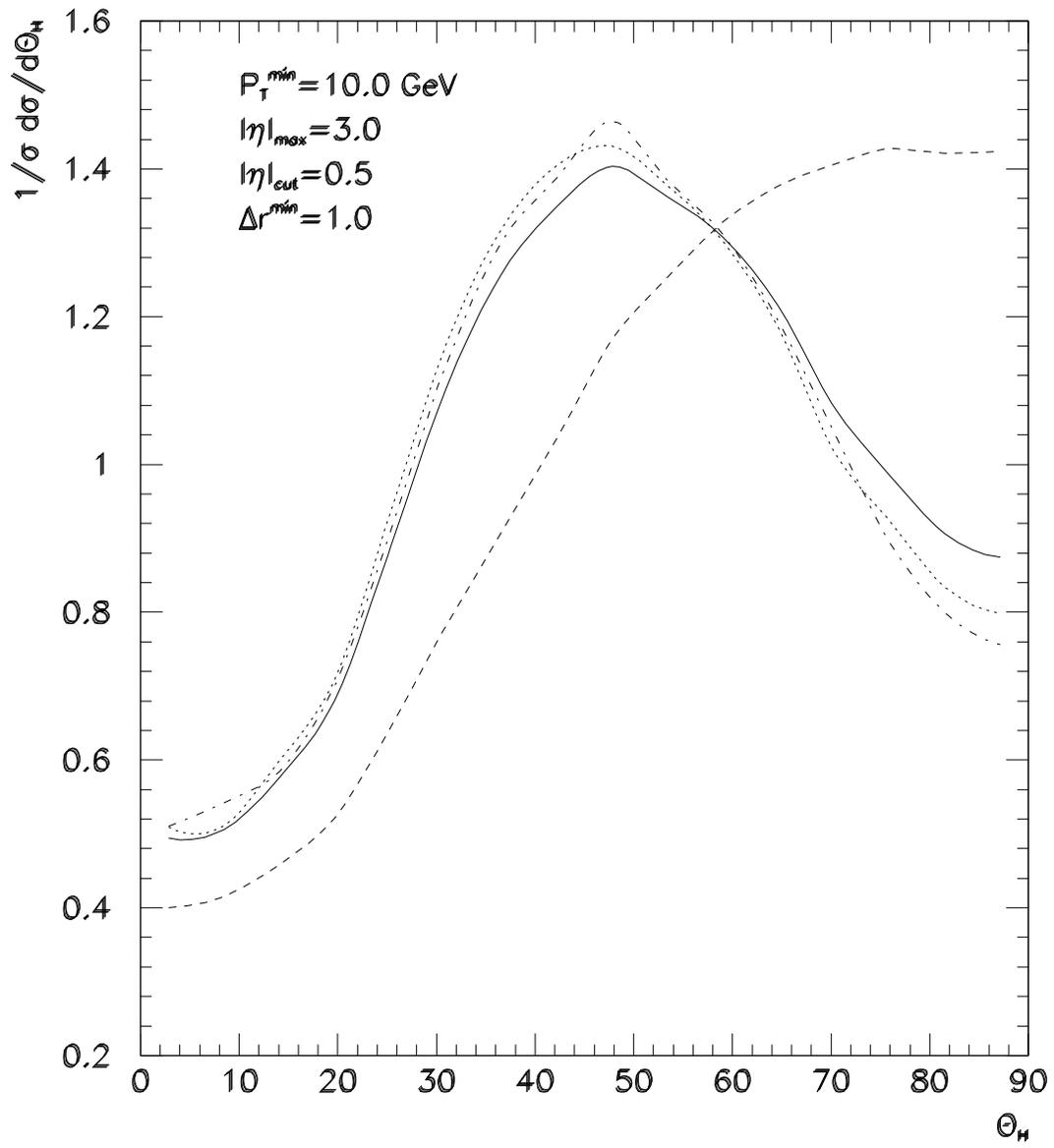
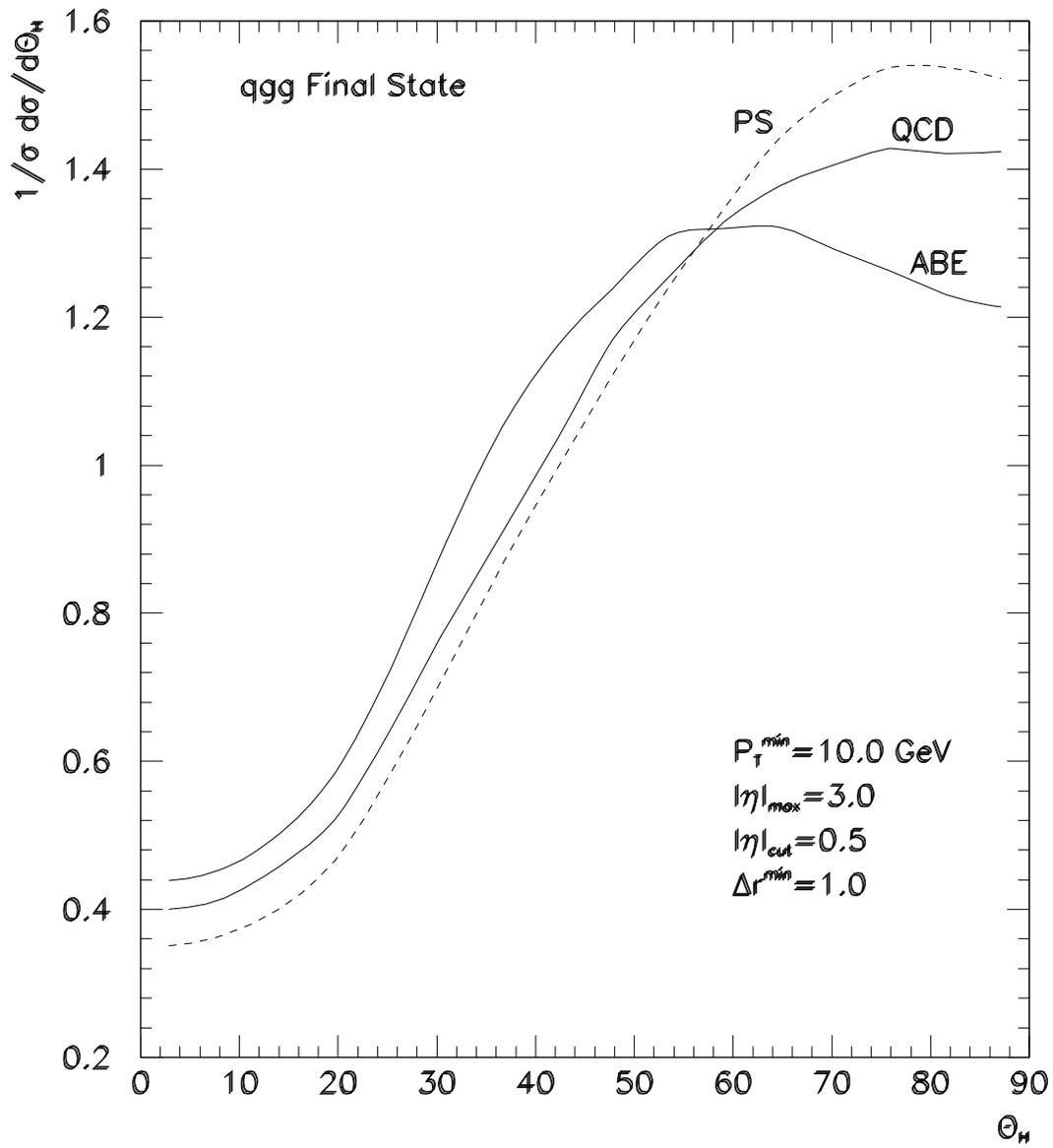
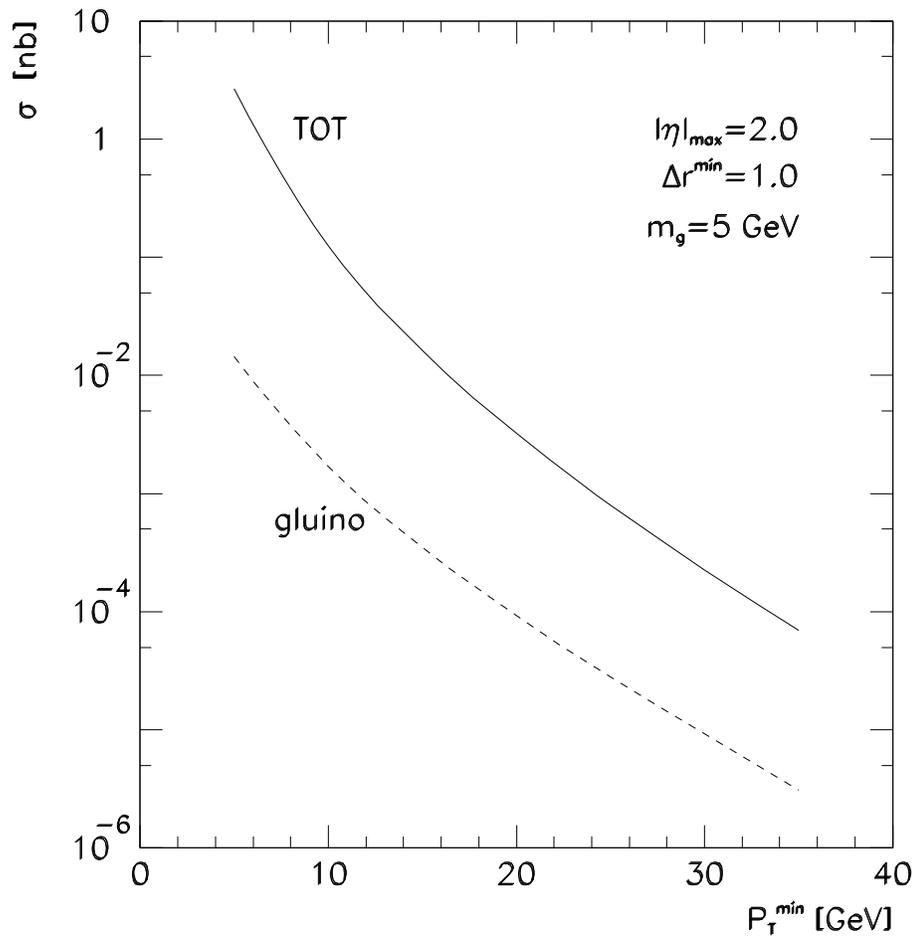
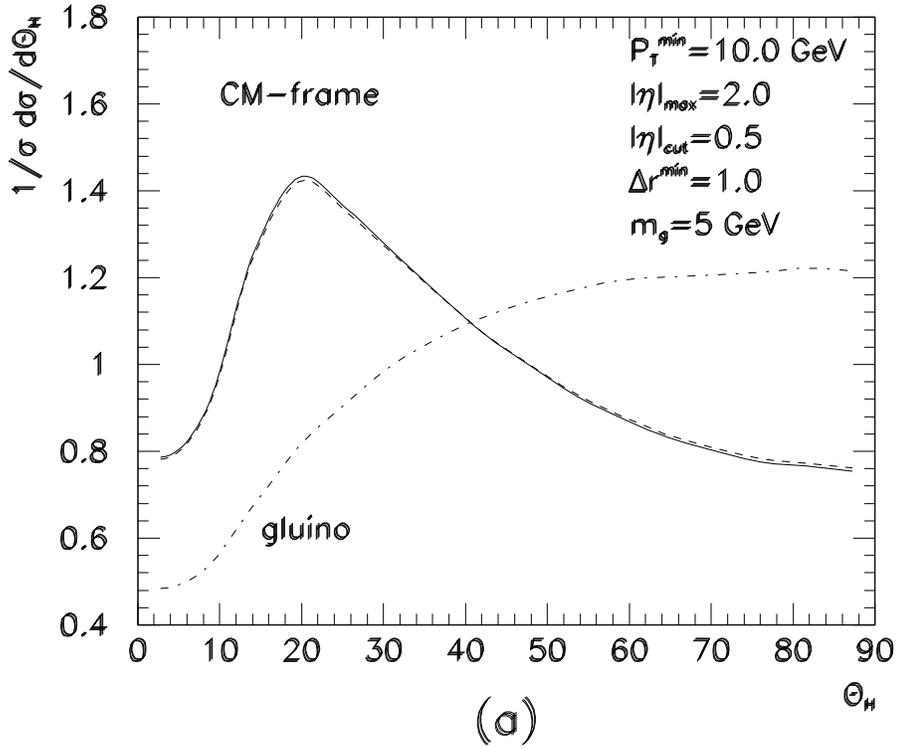
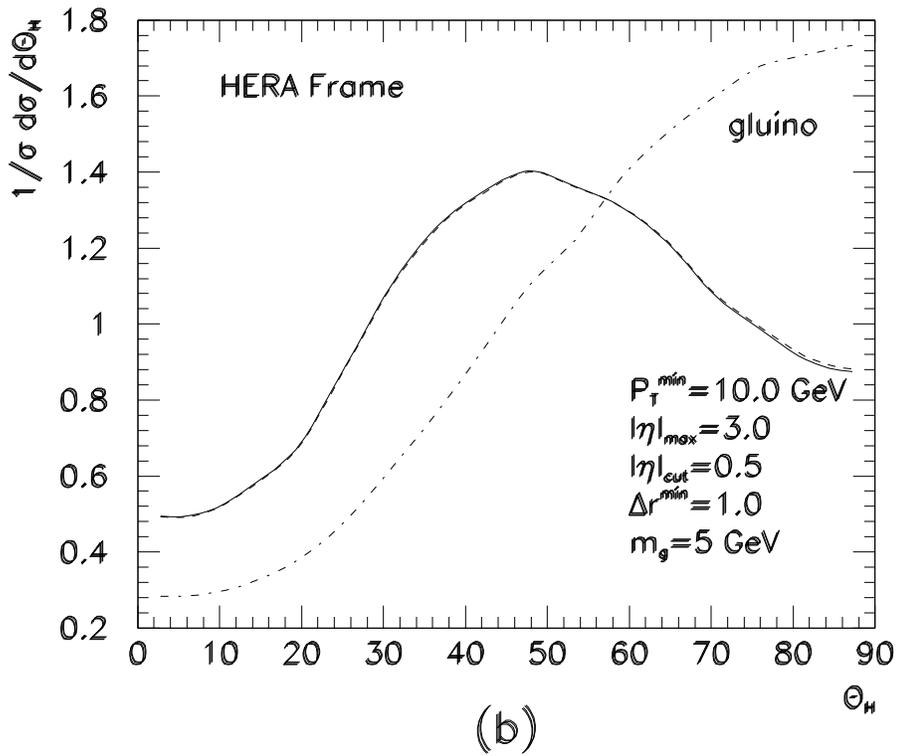
\end{document}